\documentclass[aps,prb,reprint,twocolumn,amsmath,amsfonts, amssymb]{revtex4-1}
\usepackage{graphicx}
\usepackage{bm}
\usepackage{color}
\usepackage{hyperref}

\begin{document}

%-------------------------------------

\title{Geminal replacement models based on AGP}

\author{Rishab Dutta}
\affiliation{Department of Chemistry, Rice University, Houston, TX 77005}

\author{Thomas M. Henderson}
\affiliation{Department of Chemistry, Rice University, Houston, TX 77005}
\affiliation{Department of Physics and Astronomy, Rice University, Houston, TX 77005}

\author{Gustavo E. Scuseria}
\affiliation{Department of Chemistry, Rice University, Houston, TX 77005}
\affiliation{Department of Physics and Astronomy, Rice University, Houston, TX 77005}

%-------------------------------------

\begin{abstract}
The antisymmetrized geminal power (AGP) wavefunction has a long history and is known by different names in various chemical and physical problems. There has been recent interest in using AGP as a starting point for strongly correlated electrons. 
Here, we show that in a seniority-conserving regime, different AGP based correlator representations based on generators of the algebra, killing operators, and geminal replacement operators are all equivalent. 
We implement one representation that uses number operators as correlators and has linearly independent curvilinear metrics to distinguish the regions of Hilbert space. This correlation method called $J$-CI, provides excellent accuracy in energies when applied to the pairing Hamiltonian. 
\end{abstract}

%-------------------------------------
\maketitle
%-------------------------------------

\section{Introduction}

Single reference methods have been one of the popular choices for simulating correlated electronic structure. These methods usually choose a single Slater determinant as their starting point and then add particle-hole excitations to describe correlation.
It is well known that traditional single reference methods do not provide the correct description of strongly correlated systems and often fail catastrophically when the mean-field reference determinant is restricted to respect the symmetries of the Hamiltonian \cite{Stein2014,Bulik2015, PoST2016}. 
To overcome the inadequacies of the single Slater determinant, a more sophisticated reference is needed. One way to devise a better starting point is to break one or more symmetries of the system and project them later to recover the physical part \cite{RingBook,Schmid2004,PQT2011,PHF2012}.

While for many problems in chemistry and physics the relevant symmetry-projected methods are based on spin, there are other problems where number-projection is more appropriate. We wish to use one such wavefunction, the number projected \cite{PBCS1964,Braun1998}
Bardeen-Cooper-Schrieffer (BCS) \cite{BCS1957} state, as our starting point. Number-projected BCS is equivalent to the antisymmetrized geminal power (AGP) wavefunction \cite{Coleman1965}, a product state of identical two-electron building blocks known as 
geminals \cite{Coleman1963,Surjan1999}. 
Although introduced in chemistry decades ago, AGP has largely been abandoned in chemical applications.  However, in the last ten years, there has been a renewed interest in AGP \cite{PQT2011,Neuscamman2012,Khamoshi2019} 
and AGP based methods \cite{Neuscamman2013,Sorella2014,Uemura2015,
Tsuchimochi2015,Kawasaki2016,Dukelsky2019, Uemura2019,TomAGPCI2019,TomAGPRPA2020} for accurate energy calculations.
AGP has also been used to describe thermal states \cite{Harsha2020} and implemented in near-term quantum computers \citep{AGPQC2020}.
Recent work by two of the present authors has shown that AGP is a fruitful starting point for the description of strong pairing correlations \cite{TomAGPCI2019,TomAGPRPA2020}. 
In this work, we introduce new correlated models based on AGP and show that seemingly different post-AGP models are equivalent in the sense that they all can be written in a geminal replacement representation. 
The concept of geminal replacement is extremely useful for chemistry where different electron pairs are best described by different geminals \cite{Pople1953}. Indeed, a geminal model more suitable for different electron pairs than AGP is the antisymmetrized product of interacting 
geminals (APIG) \cite{APIG1971,AyersCTC2013,AyersJCTC2013}. 
Our goal is to use AGP as a starting point to reach the computationally complex APIG state.
 
AGP conserves seniority \cite{Racah1943,Seniority2011}, which means it does not break electron pairs. We will only discuss seniority-conserving wavefunctions and systems in this article but it should be noted that AGP is a reasonable reference for seniority-breaking systems too \cite{Neuscamman2013,Sorella2014}. In other words, AGP provides an initial approximation to the seniority-zero sector \cite{Jacob2018} of a generic wavefunction. The description of residual pair-pair correlations can be achieved by a suitable choice of correlator acting on AGP.

In section~\ref{2.0}, we discuss geminals, geminal based models, and the pairing model Hamiltonian. Section~\ref{3.0} discusses several post-AGP models and presents numerical results.
In section~\ref{4.0}, we show how these various models can be described in the language of geminal replacement.

%-------------------------------------

\section{Background} \label{2.0}

To set the stage for adding correlations to AGP, we first need to describe AGP itself.  And as AGP is a geminal state, we will begin with a discussion of geminals. We will also discuss the model Hamiltonian used for all the numerical results here.

\subsection{AGP}

A geminal is simply a two-electron wavefunction and can be written in terms of a geminal creation operator 
\begin{equation} 
\Gamma^\dagger 
= \sum_{pq} \: \eta_{pq} \: c_{p}^\dagger c_{q}^\dagger,
\end{equation}
where $p$ and $q$ represent spin-orbitals and $\eta$ is their amplitude matrix. In the natural orbital representation of the geminal, the anti-symmetric 
$\eta$ matrix is transformed to a block diagonal form \cite{Hua1944,TomAGPRPA2020} and the geminal creation operator reduces to
\begin{equation} \label{eq:gem}
\Gamma^\dagger 
= \sum_p \: \eta_p \: P_p^\dagger,
\end{equation}
where the pair creation operator is 
\begin{equation}
P_p^\dagger = c_p^\dagger \: c_{\bar{p}}^\dagger.
\end{equation}
Here spin-orbital $\bar{p}$ is ``paired'' with orbital $p$. The pairing does not necessarily have to be between the $\uparrow$ and 
$\downarrow$ spins of the shared spatial orbital $p$, but is defined according to the orbital-pairing scheme of the natural orbital basis. The pair creation operator conserves seniority just like the pair annihilation and number operators
\begin{subequations}
\begin{align}
P_p &= c_{\bar{p}} \: c_p,
\\
N_p &= c_p^\dagger \: c_p 
+ c_{\bar{p}}^\dagger \: c_{\bar{p}},
\end{align}
\end{subequations}
and their commutation relations follow an su(2) algebra
\begin{subequations}
\begin{align}
[P_p, P_q^\dagger] 
&= \delta_{pq} \: ( 1 - N_p ),
\\
[N_p, P_q^\dagger] 
&= 2 \: \delta_{pq} \: P_q^\dagger.
\end{align}
\end{subequations}
Note that, the mapping of these generators to fermion pairs guarantees their nilpotency, i.e., 
$(P_p^\dagger)^2 = 0$.

Geminals are two-electron building blocks and can be used to construct a many-body wavefunction. One example of a geminal 
$n$-pair wavefunction is the aforementioned APIG,
\begin{equation} \label{eq:apig}
|\mbox{APIG}\rangle
= \Gamma_1^\dagger ... \Gamma_n^\dagger \: |-\rangle,
\end{equation}
where $|-\rangle$ is the physical vacuum and 
\begin{equation} \label{eq:ig}
\Gamma_\mu^\dagger
= \sum_p \: \eta_p^\mu \: P_p^\dagger.
\end{equation}
APIG is a variationally and conceptually powerful wavefunction but its computational cost for general Hamiltonians is combinatorial since its matrix elements lead to permanents \citep{AyersCTC2013,AyersJCTC2013}. Instead, we focus here on the AGP wavefunction where all the geminals are identical and use it as the basis for geminal replacement models eventually leading to APIG.

The AGP wavefunction of $n$ pairs is the product of $n$ identical geminals
\begin{subequations} \label{eq:agp}
\begin{align}
|n\rangle
&= \frac{1}{n!} \: \big( \Gamma^\dagger )^n \: |-\rangle
\\
&= \sum_{p_1 < ... < p_n} \eta_{p_1} ... \eta_{p_n} \: P_{p_1}^\dagger ... P_{p_n}^\dagger \: |-\rangle.
\end{align}
\end{subequations}
Thus, AGP approximates the doubly occupied configuration interaction (DOCI) \cite{DOCI1967,Seniority2011} wavefunction,
\begin{equation} \label{eq:doci}
|\mbox{DOCI}\rangle
= \sum_{p_1 < ... < p_n} D_{p_1 ... p_n} \: 
P_{p_1}^\dagger ... P_{p_n}^\dagger \: |-\rangle,
\end{equation}
the most general possible seniority-zero state, by a simple factorization of the tensor amplitude, as can be readily seen by comparing eqs.~\eqref{eq:agp} and~\eqref{eq:doci}.
AGP is variationally superior to Hartree-Fock since the latter is a special case of AGP, and because AGP is number projected BCS, it can be optimized with a mean-field cost. The product structure and low cost of AGP make it a potentially useful starting point for more sophisticated methods.

\subsection{Pairing Hamiltonian}

All of our numerical results concern the pairing Hamiltonian 
\begin{equation}
H = \sum_p \epsilon_p \: N_p - G \: \sum_{pq} \: P_p^\dagger P_q,
\end{equation}
where $p$ and $q$ represent levels. 
Due to the nilpotency of the operator $P_p^\dagger$, each level can be occupied by only one pair. Here $\epsilon_p = p$ and the interaction $G$ is associated with pair hopping between any two levels; the interaction may be repulsive ($G < 0$) or attractive ($G > 0$). Even though the pairing Hamiltonian is simplistic, it facilitates interesting physics in the attractive interaction regime, where Hartree-Fock instability towards a number-broken BCS state is observed \cite{BCSCC2014}.

Since the pairing Hamiltonian is seniority-conserving, the exact ground state is the same as the DOCI wavefunction. Instead of diagonalizing the Hamiltonian in the DOCI space, it can be solved exactly using a set of nonlinear equations \cite{Richardson1963,Richardson1964,RG2004} instead. This provides us exact energies and eigenstates of the pairing Hamiltonian, even for fairly large systems. The ground state of the pairing Hamiltonian is an APIG with the geminal coefficient
\begin{equation}
\eta_p^\mu
= \frac{1}{2 \: \epsilon_p - R_{\mu}},
\end{equation}
where $R_{\mu}$ is called the pair energy. 
The pairing Hamiltonian is part of a family of exactly solvable Hamiltonians called the Richardson-Gaudin models with special APIG wavefunctions as their ground state \cite{RG2004}.

We are interested in the pairing Hamiltonian primarily because many conventional quantum chemical methods are unable to describe its physics in the strongly attractive regime \cite{pECCD2015,PoST2016}, where superconductivity emerges. Coupled cluster methods even fail to yield real-valued energies after a certain positive $G$ value \cite{BCSCC2014}. 
It is well known that symmetry adapted coupled cluster methods fail to describe strongly correlated molecules \cite{Bulik2015} and repulsive models like the Hubbard Hamiltonian \cite{Stein2014}, perhaps due to a poor description of pairing inteactions at strong correlation \cite{Jacob2018,BruCC2014}.
While many methods struggle to describe the physics of the attractive pairing Hamiltonian, AGP captures its basic behavior reasonably well \cite{PoST2016}. Indeed, at extremely large positive $G$ values when the two-body part of the Hamiltonian is dominant, extreme AGP \cite{Coleman1965} (identical $\eta_p$) is the exact ground state eigenfunction of the pairing Hamiltonian. 
 
\subsection{Reduced density matrices}

One of the advantages of AGP as a reference wavefunction is that its expectation values are easily computed.
We define AGP reduced density matrices (RDMs) in the form $Z_{pqr...} = \langle P_p^\dagger ... N_q ... P_r \rangle$, for example
\begin{subequations}
\begin{align}
Z_{p}^{1, 1}
&= \langle N_p \rangle,
\\
Z_{pq}^{0, 2}
&= \langle P_p^\dagger P_q \rangle,
\\
Z_{pq}^{2, 2}
&= \langle N_p N_q \rangle,
\\
Z_{pqr}^{1, 3}
&= \langle P_p^\dagger N_q P_r \rangle,
\end{align}
\end{subequations}
where $\langle \: ... \: \rangle$ is short for 
$\langle n| \: ... \: |n \rangle$. The RDMs can be evaluated in terms of elementary symmetric polynomials \cite{SumESP1974,Khamoshi2019} or by number projection of BCS density matrices \cite{BCSRDM2018}. But the most efficient way to construct a RDM tensor is to use the reconstruction formulae \cite{Khamoshi2019}, which enable us to write higher-order AGP density matrices as linear combinations of lower-order density matrices, provided 
\begin{equation}
\eta_p^2 \neq \eta_q^2 \quad ( p \neq q )
\end{equation} 
is true. The reconstruction formulae can be used to compute a $k$-index RDM tensor in $\mathcal{O} (m^k)$ time, where $m$ is the number of spatial orbitals or the number of levels in the pairing Hamiltonian.

%-------------------------------
\begin{figure}[t]
\includegraphics[width=\columnwidth]{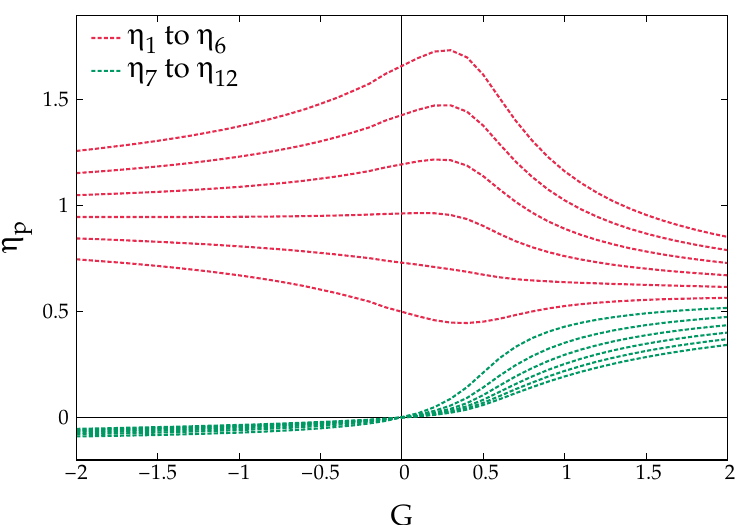}
\caption{
AGP coefficients ($\eta_p$) of a normalized AGP ($\langle n | n \rangle = 1$) for the half-filled 12-level pairing Hamiltonian. Note that AGP expectation values are invariant to a global sign change of all 
$\eta_p$.
\label{fig:etas}}
\end{figure}
%-------------------------------

Figure~\ref{fig:etas} shows the values of geminal coefficients of AGP for the pairing Hamiltonian, which has only positive values for attractive but both positive and negative values for repulsive interactions. At $G=0$, when Hartree-Fock is the ground state of the pairing Hamiltonian, $\eta_p$ corresponding to the virtual orbitals will go to zero. Also at extremely large $G$ values, the $\eta_p$ coefficients will slowly approach the same value. In other words, we can safely assume all the $\eta_p$ coefficients are different in our computations, as long as $G \neq 0$, which allows us to use the reconstruction formulae.

%-------------------------------------

\section{Correlation on AGP} \label{3.0}

Here we will discuss configuration interaction models based on AGP using killer adjoint and number operators. For the sake of simplicity, we only consider real-valued coefficients. 

\subsection{Number operator correlators}

Particle-hole excitations create a manifold of states orthogonal to the reference Slater determinant since their adjoints annihilate it; de-excitations, in other words, are killing operators of the reference determinant.  AGP also has killing operators \cite{Killer1983,TomAGPCI2019}. The seniority-conserving killing operator is
\begin{align}
K_{pq} &= \eta_p^2 \, P_p^\dagger \, P_q + \eta_q^2 \, P_q^\dagger \, P_p
\\
&+ \frac{1}{2} \, \eta_p \, \eta_q \, \left(N_p \, N_q - N_p - N_q\right),
\nonumber
\end{align}
where $p \neq q$. Because $K_{pq}$ annihilates AGP, its adjoint $K_{pq}^\dagger$ creates a manifold of states orthogonal to AGP. This leads to an AGP based configuration interaction (CI) \cite{TomAGPCI2019},
\begin{subequations}
\begin{align}
|\mbox{K-CI}\rangle 
&= ( 1 + K_2) \: |n\rangle,
\\
K_2 
&= \sum_{p > q} C_{pq} \: 
K_{pq}^\dagger,
\end{align}
\end{subequations}
where $C_{pq}$ is symmetric and intermediate normalization is assumed for the above wavefunction,
\begin{subequations}
\begin{align}
\langle n | n \rangle 
&= 1,
\\
\langle n | \: K_{pq}^\dagger \: | n \rangle 
&= 0.
\end{align}
\end{subequations}

%-------------------------------
\begin{figure*}[t]
\includegraphics[width=\columnwidth]{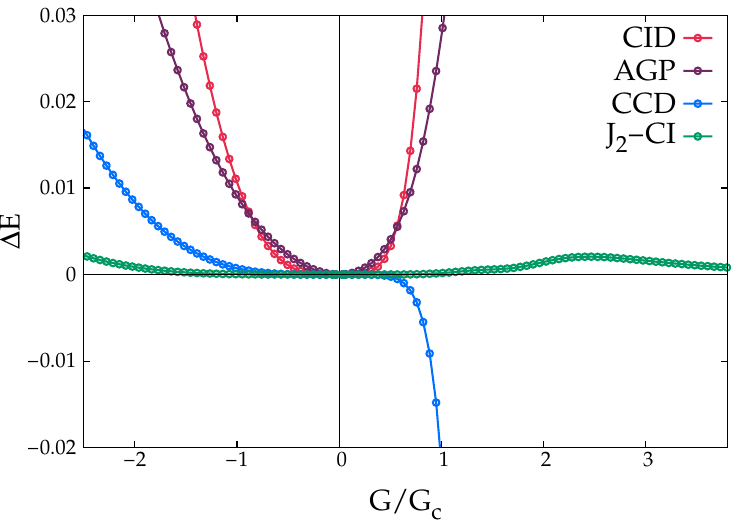}
\hfill
\includegraphics[width=\columnwidth]{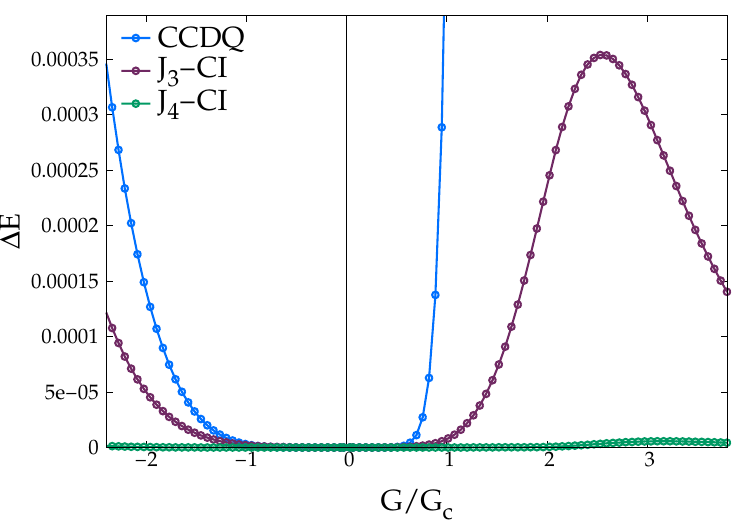}
\caption{
Total energy errors $(E_{method} - E_{exact})$ for the half-filled 12-level pairing Hamiltonian. CID, CCD, and CCDQ are post-Hartree-Fock methods: they are configuration interaction doubles, coupled cluster doubles, and coupled cluster doubles and quadruples respectively.
Note that the critical $G$ value is $G_c \sim 0.3161$.
\label{fig:en}}
\end{figure*}
%-------------------------------

Here, we formulate an alternative AGP-CI model using (Hermitian) number operator correlators, sometimes known as the Hilbert space Jastrow operators \cite{Neuscamman2013}, in the form
\begin{equation} \label{eq:jciN}
|J_k \mbox{-CI} \rangle
= \sum_{p_1 < ... < p_k} S_{p_1 ... p_k} \: 
N_{p_1} ... N_{p_k} \: |n\rangle.
\end{equation}
The symmetric amplitude tensor \textbf{S} is optimized variationally, leading to a generalized eigenvalue problem,
\begin{equation}
\mathbf{H \: S  = M \: S \: E},
\end{equation}
where for example,
\begin{subequations}
\begin{align}
H_{pq, rs}
&= \langle N_p N_q \: H \: N_r N_s \rangle,
\\
M_{pq, rs}
&= \langle N_p N_q \: N_r N_s \rangle,
\end{align}
\end{subequations}
in the case of $J_2$-CI.
Using the differential representation of $N_p$ on AGP \cite{Khamoshi2019}, 
\begin{equation} \label{eq:diff1}
N_p \: |n\rangle 
= 2 \: \eta_p \: P_p^\dagger \: |n-1\rangle,
\end{equation}
and nilpotency of the pair creation operators,
we can also write $J_k$-CI in terms of pair creation operators,
\begin{equation} \label{eq:jciP}
|J_k \mbox{-CI} \rangle
= \sum_{p_1 ... p_k} 
\tilde{S}_{p_1 ... p_k} \: P_{p_1}^\dagger ... P_{p_k}^\dagger \: |n-k\rangle,
\end{equation}
where 
\begin{equation}
\tilde{S}_{p_1 ... p_k}
= \frac{(2)^k}{k!} \: S_{p_1 ... p_k} \: ( \eta_{p_1} ... \eta_{p_k} ).
\end{equation}
Although eq.~\eqref{eq:jciN} is better suited for the computation of observables, 
eq.~\eqref{eq:jciP} helps to realize some important points, to be discussed later.

The simplest $J$-CI wavefunction has one number operator
\begin{equation}
|J_1 \mbox{-CI} \rangle
= \sum_p \: S_p \: N_p \: |n\rangle,
\end{equation}
and is not particularly interesting for ground state since it produces no correlation when acting on an optimized AGP state.  
Figure~\ref{fig:en} compares total energy errors of $J$-CI, AGP, and post-Hartree-Fock methods like configuration interaction doubles (CID), coupled cluster doubles (CCD), and coupled cluster doubles and quadruples (CCDQ) \cite{BartlettBook}, for the pairing Hamiltonian. $J_2$-CI provides far better ground state energies for the pairing Hamiltonian than CID, CCD, and AGP, both for the attractive and repulsive interactions in the strong correlation regime. The accuracy can be systematically improved with higher-order $J$-CI methods. It is clear that even $J_3$-CI performs better than the CCDQ method.

Interestingly, $J_2$-CI and K-CI yield identical energies for the pairing Hamiltonian which calls for a comparison between these two models. 
Both K-CI and $J_2$-CI have curvilinear metrics but unlike $K_2$, $J_2$ does not create correlated states orthogonal to AGP. 
In fact, the $k$-th order $J_k$-CI contains the AGP state and all lower-order $J$-CI states. 
It should be noted that although the K-CI correlator $(1 + K_2)$ adds AGP to the orthogonal manifold, it generates the same number of states as $J_2$-CI since the K-CI metric always contains one zero mode whereas the $J_2$-CI metric is positive definite. 
The higher order $J$-CI metrics are also positive definite but have near-zero modes near the Hartree-Fock limit (e.g., $G \rightarrow 0$). We will discuss the equivalence of K-CI and $J_2$-CI in terms of geminal replacements in section~\ref{4.0}.

%-------------------------------
\begin{table}[b]
\caption{Percentage of metric elements 
$> 10^{-6}$, for half-filled 12-level pairing Hamiltonian.}
\label{tab:met}
\centering
\begin{tabular}{llll}
\hline
$G$   & $J_2$-CI & $J_3$-CI & $J_4$-CI \\ 
\hline
-0.60 & 81  & 51  & 18 \\ 
-0.30 & 78  & 44  & 13 \\ 
0.30  & 89  & 61  & 22 \\
0.60  & 100 & 97  & 39 \\ 
\hline 
\end{tabular}
\end{table}
%-------------------------------

\begin{figure*}[t]
\includegraphics[width=\columnwidth]{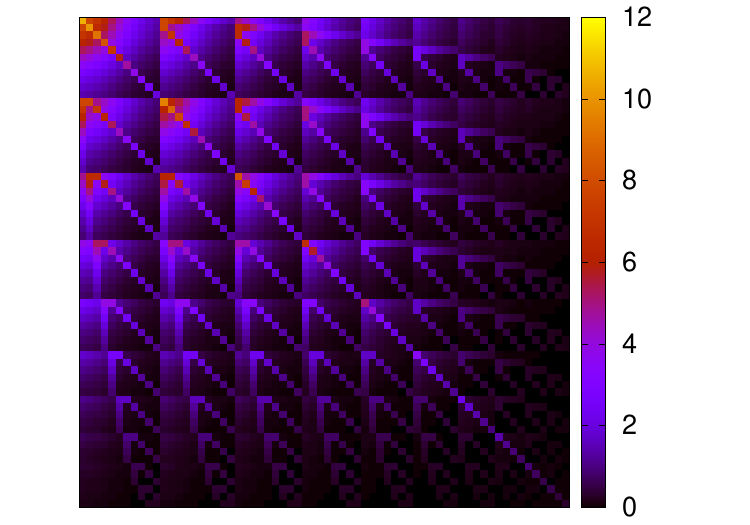}
\hfill
\includegraphics[width=\columnwidth]{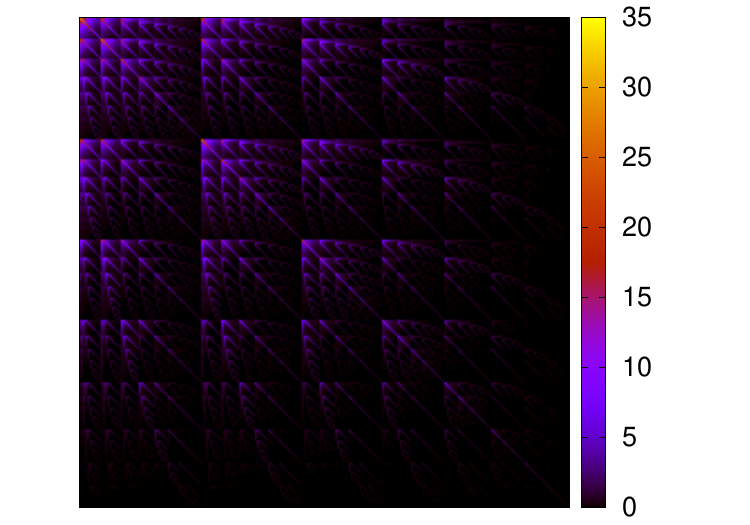}
\caption{
Metric matrix heat-maps for half-filled 12-level pairing Hamiltonian with $G = 1.20$. The left and right panels correspond to metrics of $J_2$-CI and $J_3$-CI respectively. Notice the similarity between the full matrix
on the left and the top-left section of the matrix on the right.}
\label{fig:metric}
\end{figure*}

The $J$-CI metrics are different from the Slater determinant based CI wavefunctions where the metric is the identity.
The $J_2$-CI metric is dense and although the metrics of higher order $J$-CI become less dense (Table~\ref{tab:met}), they are never the identity. In Figure~\ref{fig:metric}, we present a visualization of how metric densities change when we go from $J_2$-CI to $J_3$-CI.
It is clear from the pair creation operator representation of $J$-CI (eq.~\eqref{eq:jciP}) that the highest order $J$-CI, 
\begin{equation} \label{eq:jnci}
| J_n \mbox{-CI} \rangle 
= \sum_{p_1 ... p_n} 
\tilde{S}_{p_1 ... p_n} \: P_{p_1}^\dagger ... P_{p_n}^\dagger \: | - \rangle,
\end{equation}
is the same as DOCI but with a diagonal metric. The diagonal elements of $J_n$-CI metric
\begin{equation}
M_{p_1 ... p_n}^{p_1 ... p_n}
= \big( \eta_{p_1} ... \eta_{p_n} \big)^2 
\end{equation}
are a simple function of AGP geminal coefficients. Note that the $J$-CI metrics are non-negative, since the AGP expectation values only contain $N_p$ operators \cite{Khamoshi2019}. 

%-------------------------------
%-------------------------------

\subsection{Excited states}
Due to the nature of eigenvalue problems, we can compute excitation energies by solving $J_k$-CI methods using the difference between eigenvalues
\begin{equation}
E_{exc} = E_\mu - E_0 \quad (\mu > 0).
\end{equation}
Alternatively, we can use the Hermitian operator method (HOM) \cite{HOM1973}, which is an equation of motion \cite{Rowe1968} method for excited states tailored to a 
Hermitian correlator, e.g., the $J_k$ operators.

Because it may be unfamiliar, let us take a moment to review the HOM formalism. Consider generating exact excited states $| \mu \rangle$ by acting a Hermitian operator $Q_\mu$ on the exact ground state $|0\rangle$,
\begin{equation}
Q_\mu \: |0\rangle
= Q_\mu^\dagger \: |0\rangle
= |\mu\rangle.
\end{equation}
We apply the Schr\"{o}dinger equation
\begin{subequations} \label{eq:exc1}
\begin{align}
H \: Q_\mu \: |0\rangle 
&= E_\mu \: Q_\mu \: |0\rangle,
\\
Q_\mu \: H \: |0\rangle 
&= E_0 \: Q_\mu \: |0\rangle,
\end{align}
\end{subequations}
and take the difference to get
\begin{equation} \label{eq:exc2}
[H, Q_\mu] \: |0\rangle
= ( E_\mu - E_0 ) \: Q_\mu \: |0\rangle.
\end{equation}
Now we expand $Q_\mu$ in a Hermitian 
operator basis
\begin{equation}
Q_\mu = \sum_p \: c_p^\mu \: R_p,
\end{equation}
left-multiply eq.~\eqref{eq:exc2} by $R_p$ and take the exact ground state expectation value to arrive at
\begin{align} \label{eq:exc3}
&\sum_q \: \langle 0| \: R_p \: [H , R_q] \: 
|0\rangle \: c_q^\mu
\\
&= ( E_\mu - E_0 ) \: 
\sum_q \: \langle 0| \: R_p \: R_q \: 
|0\rangle \: c_q^\mu.
\nonumber
\end{align} 
We now take the differences of the adjoints of eq.~\eqref{eq:exc1}, right-multiply by $R_p$ and take the exact ground state expectation value to arrive at
\begin{align} \label{eq:exc4}
&-\sum_q \: \langle 0| \: [H , R_q] \: R_p \: 
|0\rangle \: c_q^\mu
\\
&= ( E_\mu - E_0 ) \: 
\sum_q \: \langle 0| \: R_q \: R_p \: 
|0\rangle \: c_q^\mu.
\nonumber
\end{align} 
Combining eqs.~\eqref{eq:exc3} and~\eqref{eq:exc4}, we get the HOM equation,
\begin{align} \label{eq:hom}
&\sum_q \: \langle 0| \:[R_p, \: [H , R_q]] \: 
|0\rangle \: c_q^\mu
\\
&= ( E_\mu - E_0 ) \: 
\sum_q \: \langle 0| \: \{ R_p, \: R_q \} \: 
|0\rangle \: c_q^\mu.
\nonumber
\end{align}
$J_k$-HOM equations are derived from above by approximating the exact ground state and excitation operator by AGP and the $J_k$ correlators respectively. For example, the $J_1$-HOM expressions are
\begin{subequations}
\begin{align}
\mathbf{H \: C}
&= \mathbf{M \: C \: \Omega},
\\
H_{pq}
&= \langle \: [N_p, \: [H, N_q]] \: \rangle,
\\
M_{pq}
&= 2 \: \langle \: N_p N_q \: \rangle,
\\
\Omega_p 
&= E_p - E_0.
\end{align}
\end{subequations}
Because of the double commutators, the resulting operator rank and RDMs are two orders lower for the $J$-HOM matrices than the corresponding $J$-CI matrices. 
 
 %-------------------------------
\begin{figure*}[t]
\includegraphics[width=\columnwidth]{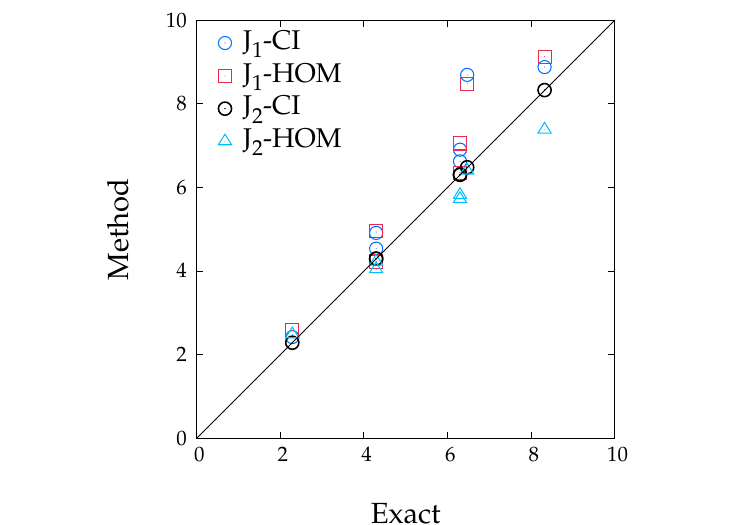}
\hfill
\includegraphics[width=\columnwidth]{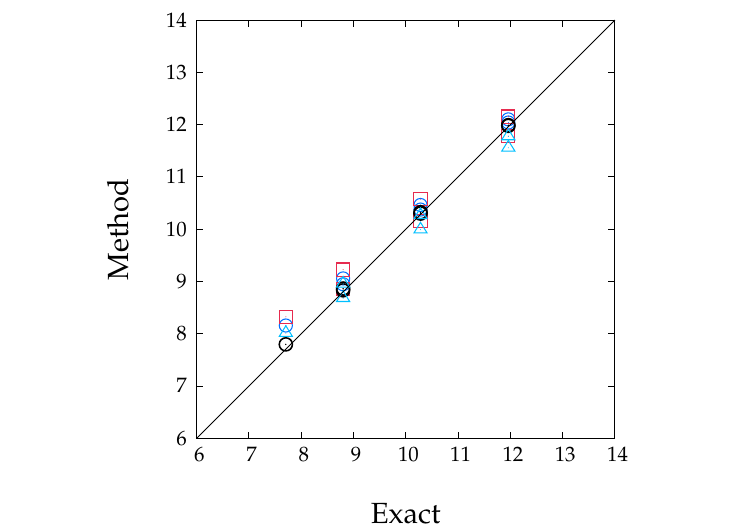}

\includegraphics[width=\columnwidth]{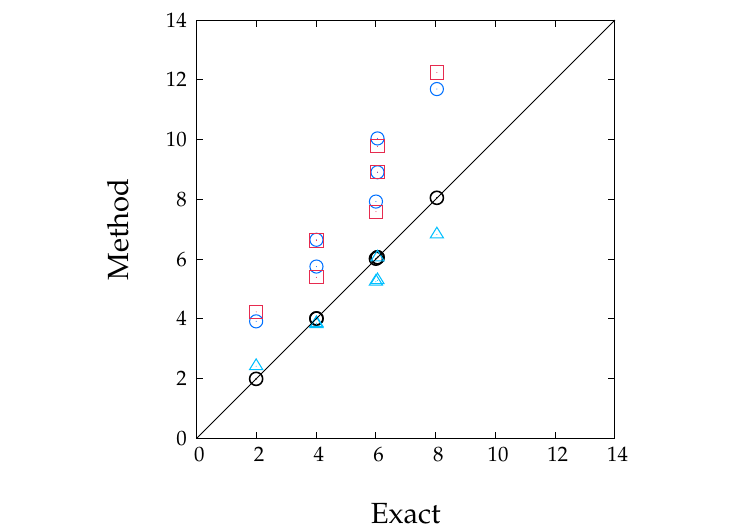}
\hfill
\includegraphics[width=\columnwidth]{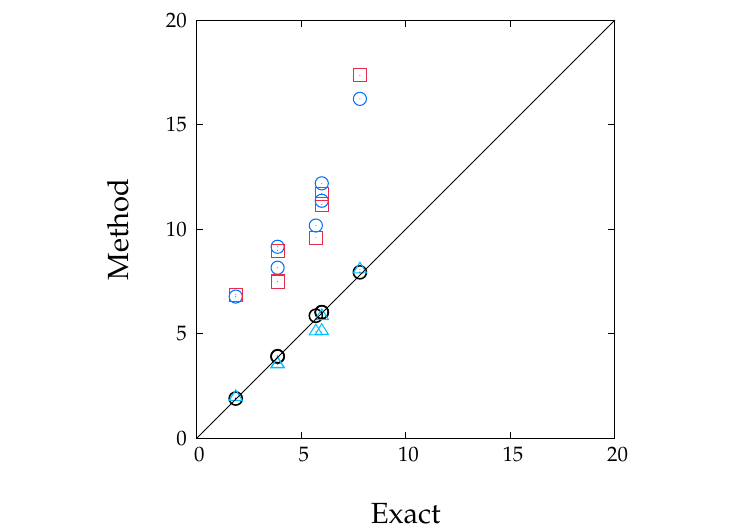}
\caption{
Excitation energies $(E_{excited} - E_{ground})$ for different $G$ values of half-filled 8-level pairing Hamiltonian. Going clockwise from top left, the $G$ values are $G=0.40$, $G=1.20$, $G=-1.20$, and $G=-0.40$ respectively. The critical $G$ value is $G_c \sim 0.3710$.}
\label{fig:exc}
\end{figure*}
%-------------------------------

Figure~\ref{fig:exc} compares the first eight seniority-conserving excitation energies computed using $J_1$-CI, $J_2$-CI, $J_1$-HOM, and $J_2$-HOM for the half-filled 8-level pairing Hamiltonian. The reason for choosing a system with 4 pairs is that the exact eigenvalue spectra can be obtained from $J_4$-CI. It is evident from Figure~\ref{fig:exc} plots that the HOM and CI excitation energies are similar for the same order of correlation. Although $J_1$-CI and $J_1$-HOM are qualitatively correct in the attractive regime, the results are far from the exact values in the repulsive regime. $J_2$-CI methods consistently perform well for all the cases shown whereas $J_2$-HOM results divert from $J_2$-CI for higher excitation energies.

%-------------------------------------

\section{Geminal replacement} \label{4.0}

In this section, we show the equivalence between different correlator representations based on generators of the algebra, killer adjoints, and geminal replacement operators.
Then we discuss a general geminal replacement model based on AGP.

\subsection{Symmetric tensor decomposition}

The $J_k$-CI (2 $\leq$ k $\leq$ n) amplitude \textbf{S} is a $k$-index $m$-dimensional symmetric tensor.
Symmetric tensors can always be decomposed \cite{Comon2008} by the symmetric form of the well-known canonical polyadic decomposition \cite{CANDECOMP1970,PARAFAC1970,Kolda2009},
so we may write
\begin{equation} \label{eq:std}
S_{p_1 ... p_k} 
= \sum_{\mu=1}^{R} \: \lambda_{\mu} \: s_{p_1}^\mu ... \: s_{p_k}^\mu, 
\end{equation}
where $R$ is the dimension of the auxiliary index $\mu$ and is called the (symmetric) rank of a tensor. Eq.~\eqref{eq:std} is also known as Waring decomposition \cite{Waring2013,Teitler2013}.
If the \textbf{s} matrix is orthogonal \cite{Anandkumar2014} then it is the natural extension of the eigen-decomposition of a symmetric matrix. Note that for a generic symmetric tensor, the rank may be too large for practical application.

Using eqs.~\eqref{eq:ig},~\eqref{eq:jciP} 
and~\eqref{eq:std}, we arrive at another representation of $J_k$-CI on AGP
\begin{equation} \label{eq:lck}
| J_k \mbox{-CI} \rangle 
= \frac{2^k}{k!} \: \sum_{\mu} \: \lambda_{\mu} \: \big( \Gamma_{\mu}^{\dagger} \big)^k \: | n-k \rangle,
\end{equation}
which writes $J_k$-CI as a linear combination of $k$-geminal replacements, where the new geminal coefficients are defined as
\begin{equation}
\eta_p^{\mu}
= \eta_p \: s_p^{\mu}.
\end{equation}
If $k=2$, the exact decomposition of $S_{pq}$ is known from the eigen-decomposition with the rank being equal to the number of levels $m$, 
\begin{subequations}
\begin{align}
S_{pq} 
&= \sum_{\mu=1}^{m} \: \lambda_{\mu} \: s_p^\mu s_q^\mu,
\\
|J_2 \mbox{-CI} \rangle 
&= 2 \: \sum_{\mu=1}^{m} \: \lambda_{\mu} \: \big( 
\Gamma_{\mu}^{\dagger} \big)^2 \: | n-2 \rangle.
\end{align}
\end{subequations}
If $k=n$, the wavefunction is
\begin{equation} \label{eq:lcagp}
| J_n \mbox{-CI} \rangle 
= \sum_{\mu=1}^{R} \: \lambda_{\mu} \: | n_\mu \rangle,
\end{equation}
where the factor $\frac{2^n}{n!}$ is absorbed into the $\lambda$ vector and the $n$-pair state $| n_\mu \rangle$ turns out to be AGP,
\begin{equation}
| n_\mu \rangle
= \big( \Gamma_{\mu}^{\dagger} \big)^n \: | - \rangle.
\end{equation}
Hence the wavefunction in 
equation~\eqref{eq:lcagp} is a linear combination of AGPs (LC-AGP) \cite{Uemura2015,Uemura2019} and is similar to the generalized BCS ansatz \cite{Egido2003}.
The symmetric tensor decomposition route to 
LC-AGP has been studied before \cite{Uemura2015,Airi2018}, but to the best of our knowledge, the natural emergence of LC-AGP from correlation on a single AGP has not been discussed before.
In principle, LC-AGP can approach the exact seniority-zero state with an increasing rank $R$. We prove in appendix~\ref{app:fis} that when $R$ is a combinatorial number $(R = 2^{n-1})$, LC-AGP is indeed equivalent to APIG which is exact for the Richardson-Gaudin models including the pairing Hamiltonian. In practice though, LC-AGP may be a numerically challenging trial wavefunction. For the pairing Hamiltonian, we have observed convergence issues and strong initial guess dependence when solving LC-AGP variationally, even for systems with 4 pairs. Nevertheless, for simple nontrivial cases like 2 pairs in 4 levels for the pairing Hamiltonian, we were able to converge LC-AGP to nearly exact answers using the expected number of terms in the expansion. Numerical issues for LC-AGP were also reported for seniority-breaking systems\cite{Uemura2015}.

\subsection{General form}

Let us discuss the simplest possible geminal operators. The geminal creation operator was defined in eq.~\eqref{eq:gem} and the AGP state is
\begin{equation} \label{eq:gem1}
|n\rangle
= \frac{1}{n} \: \Gamma^\dagger \: |n-1\rangle.
\end{equation}
We can also define a geminal removal operator 
\begin{equation} \label{eq:gem2}
\bar{\Gamma} 
= \sum_p \: \frac{1}{\eta_p} \: P_p,
\end{equation}
using the differential representation of $P_p$ on AGP \cite{Khamoshi2019},   
\begin{equation} \label{eq:diff2}
P_p^\dagger P_q \: | n \rangle
= \eta_q \: P_p^\dagger \: | n-1 \rangle
- \eta_q^2 \: P_p^\dagger P_q^\dagger \: | n-2 \rangle,
\end{equation}
which removes a geminal from AGP,
\begin{equation} \label{eq:gem3}
|n\rangle
= \frac{1}{(m-n)} \: \bar{\Gamma} \: 
|n+1\rangle.
\end{equation}
Note that $ \bar{\Gamma}^\dagger \neq \Gamma $ and eq.~\eqref{eq:diff2} reduces to 
eq.~\eqref{eq:diff1} when $p=q$.
Hence the simplest geminal replacement operator would be 
\begin{equation} \label{eq:gem4}
\Gamma_1^\dagger \: \bar{\Gamma} \: |n\rangle
= \big( \sum_p \: \eta_p^1 \: P_p^\dagger \big) \: \big( \sum_q \: \frac{1}{\eta_q} \: P_q \big) \: |n\rangle.
\end{equation}
If we only consider the diagonal part of 
the above equation, then it is the same as 
$J_1$-CI since for AGP within a seniority-conserving space, the relation
\begin{equation}
N_p = 2 \: P_p^\dagger P_p
\end{equation}
is true. If we only consider the off-diagonal part, then this is equivalent to acting with the ``pair-hopper'' operator $(P_p^\dagger P_q)$ on AGP,
\begin{equation} 
\Gamma_1^\dagger \: \bar{\Gamma} \: |n\rangle
= \sum_{p \neq q} \: \big( \frac{\eta_p^1}{\eta_q} \big) \: 
P_p^\dagger P_q \: |n\rangle,
\end{equation}
but with a factorized amplitude.

We come to an important realization. Correlators based on any generator of the algebra acting on AGP either add, remove, or replace a geminal. The number of geminals replaced by a correlator becomes more important than the nature of the correlator. This justifies why 
K-CI and $J_2$-CI provide identical energies: since $K_2$ does not contain more than two generators in each term, it carries out at most 2-geminal replacements. For the same reason, a pair-hopper based CI model
\begin{equation}
|\mbox{P-CI}\rangle
= \big( 1 + \sum_{p > q} \: t_{pq} \: 
P_p^\dagger P_q \big) \: |n\rangle,
\end{equation}
also yields identical energies to K-CI and 
$J_2$-CI. Note that both P-CI and K-CI wavefunctions add the AGP state to the excitations and both the metrics contain one zero mode.

It is natural to formulate a general geminal replacement model at this point.
We define the $k$-geminal replacement configuration interaction state as
\begin{align} \label{eq:grci}
&|k \mbox{GR-CI} \rangle 
\\
&= \sum_{\mu_1 ... \mu_k} C_{\mu_1 ... \mu_k} \: 
( \Gamma_{\mu_1}^\dagger ... \Gamma_{\mu_k}^\dagger ) \: 
( \bar{\Gamma} )^k \: | n\rangle
\nonumber
\end{align} 
or alternatively
\begin{align} 
&|k \mbox{GR-CI} \rangle 
\\
&= \sum_{\mu_1 ... \mu_k} C_{\mu_1 ... \mu_k} \: 
( \Gamma_{\mu_1}^\dagger ... \Gamma_{\mu_k}^\dagger ) \: | n-k\rangle, 
\nonumber
\end{align}
where the scalars have been absorbed into the amplitude $C_{\mu_1 ... \mu_k}$. To show the equivalence of the above wavefunction with one of the correlated models on AGP, we apply symmetric tensor decomposition of
\begin{equation} 
C_{\mu_1 ... \mu_k}
= \sum_{\sigma} \: \lambda_{\sigma} \: U_{\mu_1}^{\sigma} ... U_{\mu_k}^{\sigma},
\end{equation}
and define
\begin{align}  
&\tilde{S}_{p_1 ... p_k} 
\\
&= \sum_{\mu_1 ... \mu_k} \sum_{\sigma}
\lambda_{\sigma} \: ( U_{\mu_1}^{\sigma} ... U_{\mu_k}^{\sigma} ) \:
( \eta_{p_1}^{\mu_1} ... \eta_{p_k}^{\mu_k} ),
\nonumber
\end{align}
after we expand eq.~\eqref{eq:grci} in the Slater determinant basis, to finally get
\begin{align}
&|k \mbox{GR-CI} \rangle 
\\
&= \sum_{p_1 ... p_k} \tilde{S}_{p_1 ... p_k} \: 
( P_{p_1}^\dagger ... P_{p_k}^\dagger ) \: 
| n-k\rangle.
\nonumber
\end{align}
The above equation is nothing but the $J_k$-CI wavefunction in the pair creation operator representation (eq.~\eqref{eq:jciP}). 

%-------------------------------------

\section{Discussion} \label{5.0}

In molecular orbital based correlation theories, excitations on a reference state are achieved by replacing occupied orbitals of the reference Slater determinant with virtual orbitals. Like the one-electron molecular orbitals, geminals are the two-electron building blocks of a many-body wavefunction. It is tempting to think of correlations in geminal based models in terms of geminal replacements. But formulating a geminal replacement model with a general geminal reference is complicated. 

We have shown how to systematically build geminal replacement models starting from AGP by using the generators of the algebra or their combinations as the correlators. 
Geminal replacements are not easy to define, at least in terms of simple actions of operators, for other geminal products. This is a clear advantage of working with a basis of AGPs.
Earlier work on correlated AGP, where a killer adjoint operator creates excitations orthogonal to AGP, is also shown to be equivalent to the 
second-order correlated models using their geminal replacement representations. 
Despite the algebra being clear about the equivalence of all these representations, we have carried out numerical experiments to verify its correctness. 
The curvilinear metrics of AGP-CI wavefunctions distinguish between different regions of the DOCI space, a property not observed in the Slater determinant based CI wavefunctions, where the metrics treat all Slater determinants on an equal footing.

We have found the $J$-CI model to be the most suitable for seniority-conserving systems because of the absence of linear dependence in the metric and generalization to any order.  
$J_2$-CI provides excellent accuracy for the pairing Hamiltonian ground state and excitation energies, and adding higher order correlations systematically improves the accuracy. 
But there is room for improvement in computational efficiency. Building and diagonalizing the $J_2$-CI matrices scales reasonably (i.e., $\mathcal{O} (m^6)$) but increases exponentially with higher order $J$-CI models. The storage cost of $J_2$-CI is 
$\mathcal{O} (m^4)$, but can be reduced if an iterative diagonalization scheme is employed.
Tensor decomposition of both the RDM and amplitude tensors will be necessary to apply the higher order $J_k$-CI methods to large systems.
We have discussed the symmetric tensor decomposition of $J$-CI amplitudes (section \ref{4.0}) and decomposition of irreducible RDM tensors (section \ref{2.0}), known as the reconstruction formulae.
We are currently working on iteratively solving for the decomposed amplitudes, based on the ideas described above.
 
At the Hartree-Fock limit, $J_k$ operators will not add any correlation since any Slater determinant is an eigenfunction of the orbital number operator $N_p$. Although the pairing Hamiltonian has Hartree-Fock eigenfunctions at $G=0$, this scenario is not observed in realistic Hamiltonians. We have not seen any inconsistencies with $J_2$-CI at 
$G \rightarrow 0$ but $J_3$-CI and $J_4$-CI energies do depend on the cut-off values for the near-zero modes at small $G$ values 
($|G| \leq 0.1 \: G_c$). Note that the pair-hopper and the AGP killer adjoint correlators reduce to the traditional particle-hole excitations in the Hartree-Fock limit \cite{TomAGPCI2019}.

Some words about the correlated AGP models from a symmetry-projection point of view. 
When compared to other ideas developed in our group in the general area of combining symmetry breaking and restoration tools with correlation methods like coupled cluster theory \cite{PPS2019}, the methods presented in this article fall under the general category of project-then-correlate, as opposed to correlate-then-project, an alternative that has also been pursued both for number \cite{Ethan2019} and spin \cite{Ethan2018}.  
AGP is not size consistent and the extensive component of the AGP energy is the same as that of its underlying BCS wave function. We thus do not expect the methods discussed here to fully recover extensivity or size consistency. However, although extensivity is not well defined for the pairing Hamiltonian due to the infinite range of its interaction, we can discuss size consistency, and significant but incomplete restoration of size consistency with our post-AGP methods has been observed\cite{TomAGPRPA2020}. Note also that Neuscamman has shown that the exponential of $J_2$ acting on AGP can completely restore size consistency \cite{Neuscamman2012}.

The tools and ideas developed in this work apply strictly to seniority-conserving Hamiltonians and their eigenfunctions where all geminals, despite being different, share the same orbital-pairing scheme \cite{IG1969,AyersJCTC2013}, a property we refer to as ``coseniority'' in loose analogy to collinearity of spins.
The optimal ``different geminals for different pairs'' eigenfunctions of a seniority-breaking Hamiltonian (e.g., the molecular Hamiltonian) are bound to be noncosenior, a property indicating that different geminals have different natural orbital bases (i.e., the orbital-pairing schemes). 

Recently, Johnson \textit{et al.} have shown that the eigenfunctions of the pairing Hamiltonian can be used for the molecular Hamiltonian \cite{RG2020}. We believe that the tools and concepts developed in this work can be extended to describe seniority-breaking systems. Fundamentally, the onsite algebra of pair creation and annihilation operators 
$\{ p_{\uparrow}, p_{\downarrow} \}$ becomes one of offsite $\{ p_{\sigma}, q_{\sigma'} \}$ generators, which has a much bigger dimension, but internal structure if split into singlet and triplet components \cite{Bulik2015}. Although many details are still under development, we believe that the prospects of extending the methods presented in this paper to the molecular Hamiltonian are quite positive. Work along these lines will be reported in due time.

%-------------------------------------

\begin{acknowledgments}
This work was supported by the U.S. National Science Foundation (CHE-1762320). G.E.S. is a Welch Foundation Chair (C-0036). We thank Jorge Dukelsky for sharing his AGP code and calling our attention to the HOM method for excited states. We thank Roman Schutski for useful comments on symmetric tensor decomposition.
R.D. thanks Gaurav Harsha for valuable discussions.
\end{acknowledgments}

%-------------------------------------
\appendix
%-------------------------------------

\section{Equivalence of LC-AGP and APIG} \label{app:fis}

We express LC-AGP
\begin{equation} \label{eq:fis1}
| \Psi_1 \rangle
= \sum_{\mu} \: \lambda_{\mu} \sum_{p_1 ... p_n} (\eta_{p_1}^{\mu} \: ... \: \eta_{p_n}^{\mu}) \: 
\mathcal{P}_n^\dagger \: |-\rangle,
\end{equation}
and APIG 
\begin{equation} \label{eq:fis2}
| \Psi_2 \rangle 
= \sum_{p_1 ... p_n} 
( g_{p_1}^1 ... g_{p_n}^n ) \: 
\mathcal{P}_n^\dagger \: |- \rangle,
\end{equation}
in the Slater determinant basis, where
\begin{equation}
\mathcal{P}_n^\dagger 
= P_{p_1}^\dagger ... P_{p_n}^\dagger.
\end{equation}
Here, $n$ is the number of pairs and $g_{p}^1$ and $\eta_p^\mu$ are geminal coefficients, i.e., scalars.

Using the work of Fischer \cite{Fischer1994,Lee2016}, we can equate a product of scalars to a linear combination form
\begin{equation} \label{eq:fis3}
g_{p_1}^1 ... g_{p_n}^n 
= \frac{1}{2^{n-1} n!} \: \sum_{\mu(L) = 1}^{R} \: 
(-1)^{|L|} \: F_{p_1 ... p_n}^{\mu(L)},
\end{equation}
where the auxiliary index $\mu$ depends on the list 
$L \subset [n] = \{ 2, 3, ..., n \}$ and the length of the list is bound by $0 \leq |L| \leq n-1$. The function $F^{\mu}$ depends on the geminal matrix \textbf{g},
\begin{equation}
F_{p_1 ... p_n}^{\mu(L)} 
= \big( g_{p_1}^1 + \tau_{L, 2} \: g_{p_2}^2  + ... + \tau_{L, n} \: g_{p_n}^n \big)^n,
\end{equation}
where $\tau_{L, p} = - 1$ if $p \in L$, or 1 otherwise. We now combine eqs.~\eqref{eq:fis2} and~\eqref{eq:fis3} to get
\begin{equation}
| \Psi_2 \rangle 
= \sum_{\mu(L) = 1}^{R} \:
\frac{(-1)^{|L|}}{2^{n-1} n!} \:
\sum_{p_1 ... p_n} F_{p_1 ... p_n}^{\mu(L)} \: 
\mathcal{P}_n^\dagger \: |- \rangle.
\end{equation}
Comparing the above equation with 
eq.~\ref{eq:fis1}, we realize that linear combination of $R$ AGPs is equivalent to an APIG wavefunction if the relations
\begin{equation}
F_{p_1 ... p_n}^{\mu(L)} 
= \eta_{p_1}^{\mu(L)} \: ... \: \eta_{p_N}^{\mu(L)}
\end{equation}
and
\begin{equation}
\lambda_{\mu(L)} 
= \frac{(-1)^{|L|}}{2^{n-1} n!}
\end{equation}
are true. The total number of lists $L$ or the rank is
\begin{equation}
R = \sum_{x=0}^{n-1} \: \begin{pmatrix}
n-1 \\
x
\end{pmatrix} = 2^{n-1}.
\end{equation}
A similar discussion has also been given by
Kawasaki and Sugino \cite{Airi2018}.

%-------------------------------------
\bibliography{JCI}
%-------------------------------------

\end{document}